\documentclass[fleqn,usenatbib]{mnras}
\usepackage{newtxtext,newtxmath}
\usepackage[T1]{fontenc}
\DeclareRobustCommand{\VAN}[3]{#2}
\let\VANthebibliography\thebibliography
\def\thebibliography{\DeclareRobustCommand{\VAN}[3]{##3}\VANthebibliography}
\usepackage{graphicx,subcaption}	
\usepackage{amsmath}
\UseRawInputEncoding
\title[SED Modelling of 1ES\,1959+650]{Multiwavelength spectral and temporal analysis of VHE Blazar 1ES \,1959+650: Tracing emission mechanisms across flux states}
\author[P. Anjum et al.]{
Peer Anjum,$^{1}$\thanks{peer.anjum@iust.ac.in}
Athar A. Dar,$^{2,3}$
Zahir Shah,$^{3}$\thanks{shahzahir4@gmail.com}
Bari Maqbool,$^{1,4}$
Ranjeev Misra$^{4}$
\\
$^{1}$ Department of Physics, Islamic University of science and Technology Kashmir 192122, India\\
$^{2}$ Department of Physics, University of Kashmir, Srinagar 190006, India\\
$^{3}$ Department of Physics, Central University of Kashmir, Ganderbal 191201, India\\
$^{4}$ Inter-University Center for Astronomy and Astrophysics, Post Bag 4, Ganeshkhind, Pune, 411007, India
}

\date{}

\begin{document}
\label{firstpage}
\pagerange{\pageref{firstpage}--\pageref{lastpage}}
\maketitle

\begin{abstract}
\begin{abstract}
The high-synchrotron-peaked BL Lac object 1ES\,1959+650 exhibited pronounced activity between MJD~60310 -- 60603, including a very high energy (VHE) detection reported by LHAASO. To investigate the underlying emission mechanisms, we performed a comprehensive temporal and spectral analysis using multiwavelength data from \textit{Swift}-XRT/UVOT and \textit{Fermi}-LAT, covering the optical/UV to GeV $\gamma$-ray bands. The source shows strong energy-dependent variability, with the largest fractional variability in $\gamma$-rays, followed by X-rays and UV/optical, consistent with leptonic emission scenarios. Based on the variability patterns, we identified distinct flux states (F1, F2, F3, F4, F5, VHE-FX1, and VHE-FX2). The X-ray spectra exhibit a clear ``harder-when-brighter'' trend across these states. We modeled the broadband spectral energy distributions (SEDs) using a one-zone model incorporating synchrotron and synchrotron self-Compton (SSC) emission, implemented in \textsc{xspec} using $\chi^{2}$ minimization. During the VHE detection, the corresponding X-ray/optical emission likely resembled the F2 state. Modeling the VHE SED using F1-state data led to an SSC overprediction of the VHE flux, whereas all other states were well described within the one-zone framework. Systematic trends in physical parameters are observed across flux states, including spectral hardening, increasing break energy, rising bulk Lorentz factor, and decreasing magnetic field with increasing flux. These results suggest that enhanced particle acceleration efficiency and stronger Doppler boosting drive the observed flaring activity, while the decrease in magnetic field indicates conversion of magnetic energy into particle kinetic energy, consistent with shock-driven scenarios.
\end{abstract}


\end{abstract}




\section{Introduction}
The extra-galactic $\gamma$-ray space is dominated by blazars \citep{Dermer-Giebels-2016}, a peculiar class of radio-loud active galactic nuclei (AGNs) powered by central supermassive black hole (SMBH). Blazars consist of  strong relativistic jet pointing close to the line of sight of an observer \citep{Urry-1995}. Due to the small inclination angle, the jet emission is relativistically amplified. The non-thermal emission from blazars spans throughout the electromagnetic spectrum and exhibits a very fast variability, down to the timescale of minutes in very high energy (VHE; $\rm E_{\gamma}\geq 100$\,GeV) regime \citep{Ulrich-1997, 1996_Gaidos, Mrk_501_minute_variability, PKS_2155_minute_variability, 2020_Mrk421}. Blazars come in two flavours: Flat Spectrum Radio Quasars (FSRQs), showing strong emission/absorption line features in their optical spectra, and BL Lacertae objects (BL Lacs), which show weak or no emission/absorption line features in their optical spectra \citep{Marcha-1996, Padovani-2007, Beckmann-2012}.
\vspace{0.2em}

The broadband spectral energy distribution (SED) of blazars comprises of two peaks in $\rm \nu-\nu F_{\nu}$ space \citep{Fossati-1998, Abdo-2010}, the low-energy peak is in infrared to X-ray range, while the high energy peak is in $\gamma$-ray band, ranging from MeV-GeV. The low-energy peak is well understood by the synchrotron radiation of the relativistic electrons in the jet \citep{Blandford, Maraschi, Ghisellini-1993}. The origin of the second peak is less clear and two scenarios have been put forth to explain it. Under leptonic scenario, the high-energy peak is either attributed to the inverse Compton (IC) up-scaterring of synchrotron photons by the relativistic electrons (SSC) \citep{Konigl-1981, Ghisellini-1989, Maraschi-1992, Markus-2002} or up-scattering of photons external to the jet (External Compton; EC) \citep{begelman-1987}. These external photons can be from accretion disc \citep{Dermer-1993, Boettcher-1997}, the broad line region (BLR) \citep{Ghisellini-1996, Sikora-1994}, and the torus \citep{Blajowski-2000, Ghisellini-2008}. The broadband SED of blazars is generally well-explained by leptonic models \citep{Sahayanathan-2012, Sahayanathan_2018, Shah-2019, Shah-2021, Malik_2022, Dar-2024}, though in some cases, the observed SED is interpreted using hadronic models \citep{Muke-2003, Botcher-2013} or lepto-hadronic models \citep{Diltz-2016, Paliya-2016}. In the hadronic scenario of blazars, the high-energy emission arises from the synchrotron emission of relativistic protons \citep{Aharonian-2000, Muke-2003} or from pair cascades triggered by proton-proton or proton-photon interactions \citep{Mannheim-1993}.
\vspace{0.2em}

On the basis of location of synchrotron peak, BL Lacs are further classified as low synchrotron peaked BL\,Lacs (LBL) having peak frequency below $10^{14}$ Hz; intermediate synchotron peaked  BL Lacs (IBL) having peak between $10^{14}$ and $10^{15}$ Hz; and high synchrotron peaked BL Lacs (HBL) having peak above $10^{15}$Hz \citep{1995ApJ...444..567P}. 1ES\,1959+650 is a HBL source located at a redshift of z=0.047 \citep{Schachter-1993}. Initially the source was discovered to emit in radio band \citep{1991ApJS...75.1011G}, and its X-ray emission was later detected during the Einstein IPC Slew Survey \citep{1992ApJS...80..257E}. Several exceptional flaring events have been reported from 1ES\,1959+650 throughout the electromagnetic spectrum, including intense activity at VHE $\gamma$-rays. The first VHE $\gamma$-ray detection of 1ES\,1959+650 was made in 1998 by the Utah Seven Telescope Array \citep{Catanese_1997}. This early detection indicated the presence of VHE $\gamma$-ray emissions from this source, making it among the first extragalactic sources identified at VHE range \citep{Holder_2003}. In 2006 Magic reported further detections of the VHE $\gamma$-rays from this blazar \citep{Albert_2006}. These observations provided more data on the variability and spectral characteristics of the VHE emissions. The High Energy Stereoscopic System (H.E.S.S) has also observed 1ES\,1959+650 and reported VHE detections \citep{2005Sci...307.1938A}, contributing to the understanding of the sources high-energy spectrum and variability. The Very Energetic Radiation imagining Telescope Array system (VERITAS) has conducted extensive observations of 1ES\,1959+650, and detected VHE $\gamma$-ray emissions during various periods \citep{Aliu_2014} including a significant flaring event in 2016. These observations helped to further characterize the sources VHE-emission properties and its temporal behaviour.
Furthermore, 1ES\,1959+650 also exhibits orphan VHE flares \citep{Krawczynski_2004,Aliu_2014}, which refer to VHE $\gamma$-ray outburst that occur without a simultaneous increase in lower-energy (e.g., X-ray or optical) emission. During simultaneous multi-wavelength flaring, the X-ray and $\gamma$-ray fluxes of 1ES\,1959+650 are well correlated, typically explained by one-zone SSC models. However, orphan VHE flares challenge this interpretation, leading to the consideration of models like the hadronic synchrotron mirror model to explain such flares \citep{Böttcher_2005}. Also, the uncorrelated multi-wavelength emission can be explained by multi-zone SSC \citep{Graff_2008} or the EC process \citep{2004ApJ...601..151K}. Recently, the LHASSO collaboration announced that the blazar\,1ES1959+650 has shown dramatic VHE activity between MJD\,60347 - 60348, \citep{2024ATel16437....1X}. This suggests the blazar has been in a high state, with a monthly average flux $(\rm E > 100 MeV)$ of approximately $(1.80\, \pm\, 0.20)\times10^{-7}\rm \,ph\,cm^{-2}\,s^{-1}$. This represents a significant increase compared to the average flux reported in the 4th \emph{Fermi}-LAT source catalogue, approximately four times higher, and is similar in brightness to the flaring activity observed during 2015 and 2016. This motivated us to study this blazar source during MJD\,60310 - 60603. The primary goal of this study is to examine the behaviour of the source 1ES\,1959+650 in different energy bands. Additionally, we sought to analyse the spectral characteristics of the source during flaring events in \emph{Fermi}, \emph{Swift}-XRT and \emph{Swift}-UVOT. The manuscript is structured as follows: In \S \ref{observation}, we present the detailed data reduction procedures from different observatories. In \S \ref{temporal}, we describe the temporal study of the source and \S \ref{broadband} is dedicated to detailed broadband SED modelling of 1ES 1959+650, and in \S \ref{summary}, we discuss and summarize our main findings. A cosmology with $\Omega_{\Lambda}$ = 0.73, $\Omega_m$ = 0.27, and $H_0$= 71 Km s$^{-1}$ Mpc$^{-1}$ is used in this work.

\section {Observations and data analysis}\label{observation}
\indent
In this study, we performed a detailed multiwavelength temporal and spectral analysis of the VHE blazar 1ES\,1959+650. We utilized observations from the \emph{Fermi}-LAT, \emph{Swift}-XRT, and \emph{Swift}-UVOT, covering the period MJD\,60310 – 60603.
During this period, the source exhibited significant activity. On MJD 60347, the Large High Altitude Air Shower Observatory (LHAASO) detected a VHE $\gamma$-ray flare from 1ES 1959+650, indicating enhanced high-energy activity. The event accumulated a significance of 8.7$\sigma$ with a flux of approximately 0.5 crab units above 1 TeV, continuing until MJD\,60348 \citep{atel16437}. This VHE flare prompted a series of rapid follow-up observations across the electromagnetic spectrum. Subsequently, Target of Opportunity observations were triggered with the Neil Gehrels Swift Observatory, and a strong X-ray flare was observed on MJD\, 60352, with a count rate of $\sim 20.35\, \mathrm{counts\,s^{-1}}$ in the 0.3\,--\,10
 keV band more than twice the rate seen in prior observations. The X-ray spectrum during this flare was notably hard, suggesting the presence of high-energy particles within the relativistic jet \citep{atel16449}. Further evidence of high activity was provided by the \emph{Fermi}-LAT, which detected enhanced emission in the GeV regime between MJD 60338 and 60357, \citep{atel16456}.These coordinated, multi-instrument observations provided a comprehensive view of the blazar's variable behavior during the selected timeframe. Notably, 1ES\,1959+650 is part of the \emph{Fermi}-LAT monitored source list, which includes sources of particular interest due to their significant variability and prominence in the $\gamma$-ray sky. The \emph{Fermi}-LAT, operating in all-sky scanning mode, provides continuous monitoring of these sources, enabling comprehensive temporal investigations. During the period MJD 60310 -- 60603 (2024 January 1 -- 2024 October 20), Swift provided pointed observations, often triggered by activity detected by Fermi. We included all available observations from Swift-UVOT/XRT and Fermi-LAT within this timeframe. Additionally, we incorporated VHE $\gamma$-ray data in our spectral analysis, where 1ES 1959+650 was detected at flux levels corresponding to a few percent of the Crab Nebula flux.

 \indent
 \subsection{\emph{FERMI}-LAT}

\indent
The \emph{Fermi} large area Telescope (LAT) is a pair conversion telescope designed to cover an energy range from 20 MeV to over 300 GeV \citep{2009ApJ...697.1071A}. It is the product of a global collaboration involving NASA and the DOE in the United States, along with various research institutions in France, Italy, Japan, and Sweden. \emph{Fermi} scans the entire sky every 3 hours in its normal scanning mode. 
Here, we have analyzed the LAT data of 1ES\,1959+650 during the period MJD\,54682 - 60603.
To prepare this data for scientific analysis, we processed it using FERMITOOL1-v2.2.0 software, following the standard analysis procedures outlined in the \emph{Fermi}-LAT documentation. Specifically, We extracted P8R3 events from a $15^{\circ}$ region of interest centered on the source location, selecting events with a high probability of being photons by using SOURCE class events, with parameters set to ‘evclass = 128, evtype = 3’. Additionally, photons arriving from Zenith angles greater than $>$ $90^{\circ}$ were excluded to prevent contamination from Earth Limb  $\gamma$-rays. For spectral analysis, we considered photons in the energy range of 0.1-300 GeV. we also used the latest version of FERMIPY-v1.1.0 in our analysis. 
 The latest \textit{Fermi}-LAT 4FGL catalogue \citep{Abdollahi_2020} was employed to generate the XML model file, which includes all sources within the region of interest (ROI) along with their spectral models, positions, and normalizations. In this XML model file, the Galactic diffuse emission model \texttt{gll\_iem\_v07.fits} and the isotropic background model \texttt{iso\_P8R3\_CLEAN\_V3\_v1.txt} were utilized. The post-launch instrument response function applied was \texttt{P8R3\_SOURCE\_V3}. For the SED modeling, the flux points and energies from \textit{Fermi}-LAT observations were converted into XSPEC-readable (PHA) format using the \texttt{ftflx2xsp} tool.

 \indent
 
\subsection{\emph{SWIFT}-XRT}
 \indent
The X-ray data for our study were obtained using the \emph{Swift}-XRT instrument aboard the Neil Gehrels \emph{Swift} Observatory \citet{2004ApJ...611.1005G}. The detector has an area of $135\,\mathrm{cm}^2$ covering an energy range of 0.3\,--\,10 keV. 
Data from the \emph{Swift}-XRT instruments are available through NASA's HEASARC archive. The \emph{Swift} observatory conducted 22 observations of 1ES\,1959+650 (see Table \ref{obse_id}) during the period from MJD\,60310 - 60603. Each \emph{Swift} observation ID corresponds to a single data point in the X-ray light curve. We processed the X-ray data collected in the photon-counting mode using the XRTDAS v3.7.0 software package, which is part of HEASOFT package (version 6.32.1). Following the standard procedures outlined in the \emph{Swift} analysis thread, we used XRTPIPELINE (version 0.13.7) to generate the level 2 cleaned event files. Source events for the spectral analysis were selected from a circular region with a radius of 50 arcseconds, and background spectra were chosen from a circular region with a radius of 100 arcseconds. Exposure maps were aggregated using XIMAGE, and auxiliary response files were created with the task xrtmkarf. The source spectra were binned with the task GRPPHA so that each bin contained at least 20 counts. Spectral analysis was performed using XSPEC version 12.13.1 \citep{1996ASPC..101...17A}. The X-ray spectrum was fitted with a log-parabola model incorporating absorption due to neutral hydrogen (Tbabs). The column density of neutral hydrogen was fixed at $N_\mathrm{H} = 1.01 \times 10^{21}\, \mathrm{cm^{-2}}$, This value corresponds to the Galactic hydrogen column density obtained from \citep{Kalberla_2005}, while the normalization and spectral parameters of the log-parabola model were kept as free parameters.

\begin{table*}
    \centering
    \caption{Details of \emph{Swift}-XRT/UVOT observations of 1ES 1959+650 during the period MJD\, 60310 -- 60603. Column 1: Observation ID; Column 2: Date of observation; Column 3: XRT exposure time (in seconds); Column 4: UVOT exposure time (in seconds).}
    \label{obse_id}
    \begin{tabular}{cccc} 
        \hline
        Observation ID & Date & Exposure (XRT) & Exposure (UVOT) \\
        \hline
        00013906101 & 2024-02-14 & 947.20 & 916.67 \\
        00013906102 & 2024-02-15 & 839.45 & 813.18 \\
        00013906103 & 2024-02-16 & 840.62 & 813.36 \\
        00013906104 & 2024-02-17 & 914.61 & 886.29 \\
        00013906105 & 2024-02-22 & 835.64 & 805.57 \\
        00013906106 & 2024-02-24 & 897.84 & 869.26 \\
        00013906107 & 2024-02-26 & 859.84 & 832.23 \\
        00013906108 & 2024-02-28 & 781.41 & 752.80 \\
        00013906109 & 2024-03-01 & 905.21 & 877.58 \\
        00013906110 & 2024-03-06 & 959.44 & 929.90 \\
        00013906111 & 2024-03-10 & 1104.44 & 1076.38 \\
        00013906112 & 2024-03-14 & 446.60 & 418.18 \\
        00013906115 & 2024-05-03 & 932.66 & 903.24 \\
        00013906116 & 2024-05-07 & 884.45 & 858.68 \\
        00013906117 & 2024-05-11 & 961.95 & 933.61 \\
        00013906118 & 2024-05-14 & 1095.24 & 1038.27 \\
        00013906119 & 2024-07-19 & 766.29 & 737.44 \\
        00013906120 & 2024-07-23 & 894.38 & 864.05 \\
        00013906121 & 2024-07-27 & 1026.42 & 996.47 \\
        00013906122 & 2024-07-31 & 1060.82 & 1031.47 \\
        00013906123 & 2024-08-04 & 929.56 & 901.94 \\
        00013906124 & 2024-08-08 & 852.65 & 822.66 \\
        \hline
    \end{tabular}
\end{table*}

\indent
\subsection{\emph{SWIFT}-UVOT}
\indent

\emph{Swift} provides not only X-ray data but also optical/UV observations through the UVOT instrument \citep{2005SSRv..120...95R}. It conducts observations in the optical and UV bands of the electromagnetic spectrum using several filters: three optical filters (U,B,V) and three UV filters (UVW1, UVW2, UVM2) \citep{2008MNRAS.383..627P}. Data from 1ES 1959+650 observed with the UVOT telescope were processed into scientific products using the HEASOFT package (version 6.32.1). The UVOT Source task within HEASOFT was employed for image processing, and the UVOTIMSUM tool was used to combine multiple images obtained through different filters. For the extraction of source counts, 5 arcsecond circular radius was chosen as the source region, centred on the target object and 10 arcsecond radius was selected as the background region. Following \citep{2011ApJ...737..103S}, the observed flux was corrected using Galactic extinction values of ${E(B-V)}$\,=\, 0.150 and ${R_v={A_v}/{E(B-V)}}$\,=\,3.1. The flux points and energies from UVOT observations were then converted into PHA file using the $‘ftflx2xsp’$ tool.

\section{TEMPORAL ANALYSIS}\label{temporal}
Given the significant activity of 1ES\,1959+650 across multiple energy bands during the period MJD\,60310 - 60603, we initiated our temporal analysis using data from the Fermi-LAT. Its continuous, all-sky monitoring capability makes Fermi-LAT particularly well-suited for tracking the long-term variability of 1ES\, 1959+650.
To examine  the temporal behavior in detail, we extracted a one-day binned $\gamma$-ray light curve of the source using \emph{Fermi}-LAT data spanning MJD\,54682 to 60603, this extended duration encompasses a broad temporal baseline that includes both high and low flux states including recent flaring activity reported in different energy bands. 
The one-day binned $\gamma$-ray light curve
exhibits a peak integrated photon flux of $(5.48 \pm 1.90) \times 10^{-7}~\mathrm{ph\ cm^{-2}\ s^{-1}}$ around MJD 57380. This peak is approximately twenty times higher than the source’s average baseline flux of $(3.31\, \pm\, 0.25) \times 10^{-8}~\mathrm{ph\ cm^{-2}\ s^{-1}}$. The photon index during this high-flux state was found to be $(1.86 \pm 0.14)$, indicating a harder spectrum.
 Spectral hardening during flaring states is a common feature in high-frequency peaked BL Lac objects and suggests the presence of efficient particle acceleration mechanisms during these episodes.
Interestingly, during the time of reported VHE activity from the source, particularly around MJD\,60347 – 60348, the \emph{Fermi}-LAT $\gamma$-ray flux was in higher flux state and  we noted a harder spectrum with a photon index of $(1.56\,\pm\,0.17)$ compared to peak $\gamma$-ray flux. While both the GeV and VHE $\gamma$-ray fluxes show enhanced activity, the hardening of the photon index in the GeV band may reflect spectral evolution that could be attributed to either the emergence of a second emission region or changes in the underlying particle population. This scenario is further explored in the broadband SED modeling section.
\vspace{0.2em}

\begin{figure*} 
	\centering
	\includegraphics[scale=0.46]{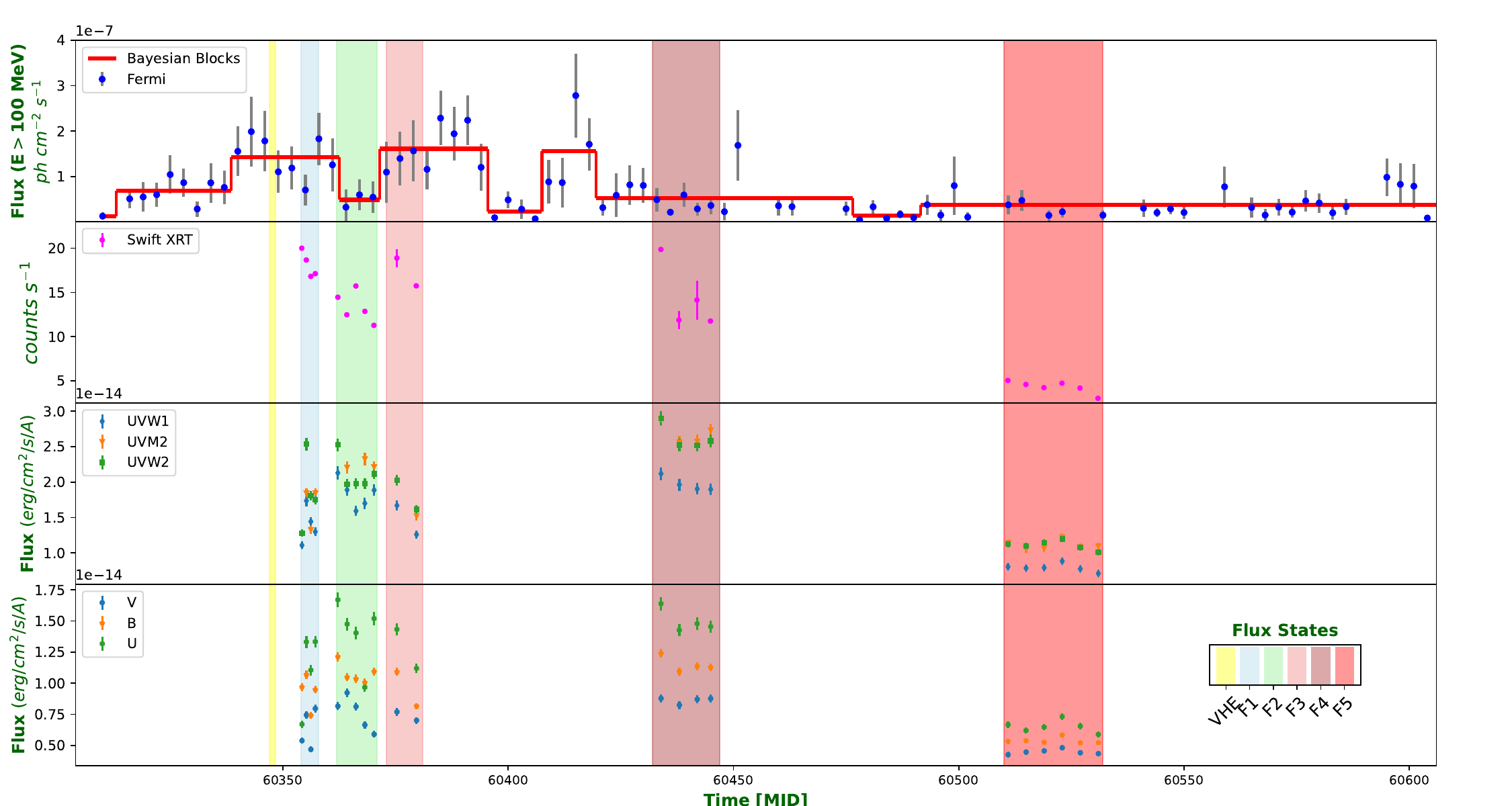}
        \caption{Multi-wavelength light curve of 1ES 1959+650 in different flux states. The top panel displays the 3-day binned light curve integrated over the energy range 0.1–300 \,GeV, including data points with $|\mathrm{TS}| > 4$, the upper middle panel displays the X-ray light curve in the energy 0.3-10\,keV, and the lower panel and the bottom panel display UV  and Optical light curves, respectively.} 
   \label{MWLC}
\end{figure*}

    To further investigate the behavior of 1ES 1959+650 during its active phase, we constructed multiwavelength light curves for the period MJD\,60310 - 60603, which are presented in Figure~\ref{MWLC}. The time window was selected based on ATel reports which reported enhanced activity across multiple energy bands, particularly in the X-ray and VHE $\gamma$-ray regimes, as well as the availability of contemporaneous multiwavelength data. The top panel  of Figure~\ref{MWLC} displays the 3-day binned $\gamma$-ray light curve derived from \emph{Fermi}-LAT observations, covering the 0.1 – 300~GeV energy range. The subsequent panels show light curves obtained from \emph{Swift}-XRT (X-ray, upper middle panel), \emph{Swift}-UVOT optical bands (lower middle panel), and UV bands (bottom panel). Each data point in the X-ray and UV/optical bands corresponds to an individual observation ID from the \emph{Swift} archive. A vertical shaded yellow band indicates the period of VHE detection (MJD\,60347 - 60348), during which the source exhibited enhanced $\gamma$-ray flux. However, the source was not observed by \emph{Swift} during this interval. Instead, X-ray observations were conducted a few days later initially capturing the tail end of the flare's decay phase, followed by additional observations during both quiescent and active states.
The multi-wavelength light curves especially in the X-ray and $\gamma$-ray bands demonstrate coordinated variability, suggesting possible correlated emission. 
During MJD 60338--60357, the 3-day binned $\gamma$-ray light curve exhibits a maximum integrated flux of $(2.78 \pm 0.09) \times 10^{-7}~\mathrm{ph~cm^{-2}~s^{-1}}$ on MJD 60415; however, no X-ray observations were available during this period. The X-ray flux peaks at $19.9~\mathrm{counts~s^{-1}}$ around MJD 60354, while the U-band flux reaches peak value at $(1.67 \pm 0.07) \times 10^{-14}~\mathrm{erg~cm^{-2}~s^{-1}~\AA^{-1}}$ and the UVW2 flux attains maximum value at $(2.90 \pm 0.05) \times 10^{-14}~\mathrm{erg~cm^{-2}~s^{-1}~\AA^{-1}}$ near MJD 60434. To examine the correlation of flux variations across different energy bands, we performed a Spearman rank correlation analysis using simultaneous $\gamma$-ray, X-ray, and optical/UV observations. The analysis reveals a strong positive correlation between the $\gamma$-ray and X-ray fluxes, with a correlation coefficient of $\rho = 0.732$ and $p$-value = 0.001, indicating a statistically significant connection between these bands. In contrast, the $\gamma$-ray flux shows weak correlations with the optical/UV bands. Details of all correlation coefficients and $p$-values are provided in  Table~\ref {tab:corelation}. These results suggest that while the $\gamma$-ray and X-ray flux variations are correlated, the non-simultaneous enhancement in the optical band may indicate distinct emission mechanisms at this band.


\begin{table}
\centering
\caption{Spearman rank correlation coefficients ($\rho$) and p-values for simultaneous multiwavelength fluxes of 1ES 1959+650. $\rho$ indicates the correlation coefficient and p-value indicates the probability of the null hypothesis.}
    \label{tab:corelation}
\begin{tabular}{lcc}
\hline
Light Curve & $\rho$ & p-value \\
\hline
$\gamma$-ray vs X-ray   & 0.732 & 0.001 \\
$\gamma$-ray vs V       & 0.017 & 0.948 \\
$\gamma$-ray vs B       & 0.199 & 0.474 \\
$\gamma$-ray vs U       & 0.217 & 0.435 \\
$\gamma$-ray vs W1      & 0.293 & 0.253 \\
$\gamma$-ray vs M2      & 0.318 & 0.288 \\
$\gamma$-ray vs W2      & 0.270 & 0.310 \\
X-ray vs V              & 0.544 & 0.013 \\
X-ray vs B              & 0.537 & 0.001 \\
X-ray vs U              & 0.484 & 0.026 \\
X-ray vs W1             & 0.544 & 0.013 \\
X-ray vs M2             & 0.637 & 0.005 \\
X-ray vs W2             & 0.542 & 0.010 \\
\hline
\end{tabular}
\label{tab:spearman}
\end{table}

Further, to quantify the variability observed in the multiwavelength light curves, we calculated the fractional variability amplitude ($F_{\mathrm{var}}$) across different energy bands. For the $F_{\mathrm{var}}$ calculation, we considered only those data points that are simultaneous across different energy bands.
This allowed us to investigate how the degree of variability depends on energy. The calculations follow the prescription outlined by \citet{2003MNRAS.345.1271V}, where the $F_{\mathrm{var}}$ is calculated using the equation:\\
\begin{equation}
\rm F_{var}=\sqrt{\frac{S^2-\overline{\sigma^{2}_{err}}}{\overline{F}^2}}
\end{equation}
where $\rm S^2$ is the variance of the flux, $\rm \overline{F}$ is the mean flux, and $\rm \overline{\sigma^2_{\mathrm{err}}}$ is the mean square of the measurement uncertainties on the flux points. The uncertainty on $F_{\mathrm{var}}$ is also computed following the method provided by \citet{2003MNRAS.345.1271V}\\
\begin{equation}
\rm F_{var,err}=\sqrt{\frac{1}{2N}\left(\frac{\overline{\sigma^{2}_{err}}}{F_{var}\overline{F}^2}\right)^2+\frac{1}{N}\frac{\overline{\sigma^{2}_{err}}}{\overline{F}^2} }
\end{equation}
\vspace{0.5em}

\begin{figure}
	\centering
	\includegraphics[scale=0.5]{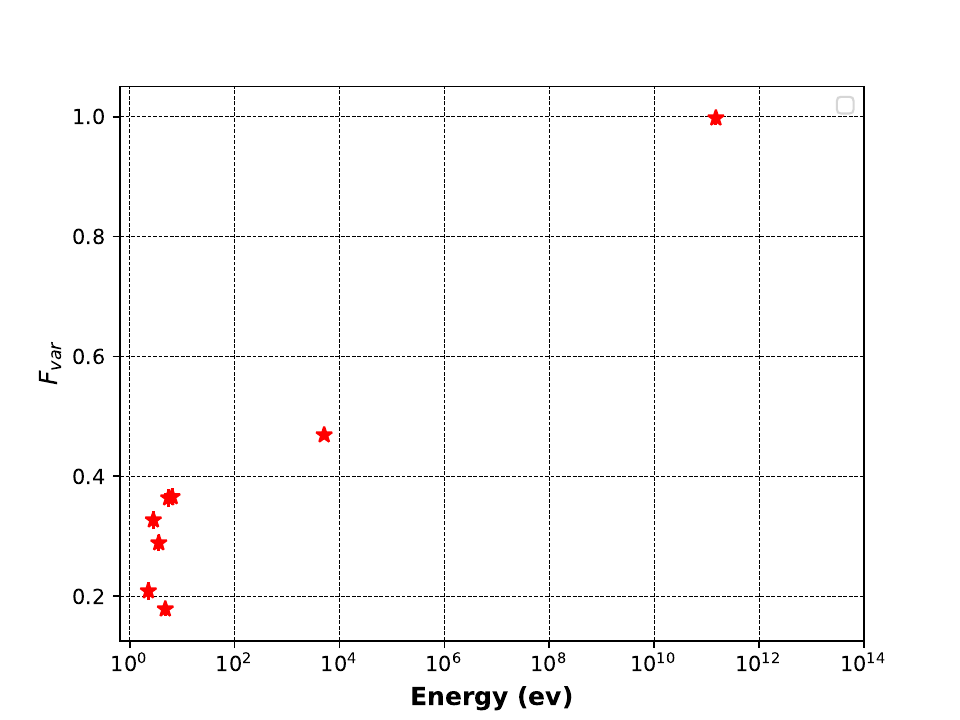}
        \caption{Energy-dependent fractional variability in different energy bands.}
    \label{frac_var}
\end{figure}

Where, N is the number of simultaneous data points in the light curve across all energy bands.  
\begin{table}
	\centering
\caption{Fractional amplitude variability of 1ES\,1959+650 in different energy bands with simultaneous data across the light curve.}
	\label{f_vart}
	\begin{tabular}{lcc} 
		\hline
		Energy Band  &  Fvar\\
		\hline
		$\gamma$-ray (0.1\,-\,300 GeV) & 0.997 $\pm$ 0.010 \\
		X-ray (0.3\,-\,10 keV) & 0.468 $\pm$ 0.005 \\
		UVW2 & 0.365 $\pm$ 0.013 \\
            UVM2 & 0.363 $\pm$ 0.015  \\
            UVW1 & 0.178 $\pm$0.010  \\
            U   & 0.288 $\pm$ 0.013  \\
            B  &  0.327 $\pm$ 0.014  \\
            V  & 0.208 $\pm$ 0.015  \\
         \hline
	\end{tabular}
\end{table}
In Figure \ref{frac_var}, we show the plot between $\rm F_{var}$ and energy, and Table \ref{f_vart} summarizes the obtained values of $\rm F_{var}$  for $\gamma$-ray, X-ray, and UVOT bands. Our results show that the $F_{\mathrm{var}}$ values increase with energy. Specifically, the lowest variability is observed in the optical/UV bands, with progressively higher values in the X-ray and $\gamma$-ray regimes. 
The $\gamma$-ray band $(0.1\text{–}300\,\mathrm{GeV})$
 shows the highest variability, with $F_{\mathrm{var}}$ = $0.99 \pm 0.01$.  Such high value of variability in the $\gamma$-ray band is sign of dynamic and energetic processes within the jet, likely driven by rapid particle acceleration. Following the $\gamma$-rays, the X-ray band $(0.3\text{–}10\,\mathrm{keV})$ exhibits the next highest $F_{\mathrm{var}}$ value at $0.46\,\pm\,0.005$. Although lower than in the $\gamma$-ray range, this level of variability is still substantial.
The increasing trend of $F_{\mathrm{var}}$ with energy suggests that the source exhibits more pronounced flux variations at higher energies. The comparatively lower variability observed in the optical/UV bands can be attributed to emission from lower-energy electrons that lie below the break energy of the particle distribution. These electrons cool more slowly and radiate from a larger and more stable region of the jet, which suppresses rapid flux changes and leads to smoother variability patterns. In contrast, the X-ray and $\gamma$-ray emissions originate from higher-energy electrons located near or above the break energy. These electrons cool efficiently through synchrotron and inverse Compton processes on shorter timescales, leading to stronger and faster variability. This energy-dependent variability is consistent with standard leptonic emission models for HBLs, where higher-energy photons are produced by higher-energy electrons that cool more rapidly, leading to faster variability. Notably, these results align with the findings of \citep{Tagliaferri_2008}, who reported that 1ES 1959+650 exhibited stronger variability in the X-ray band compared to the optical. A similar trend of increasing $F_{\mathrm{var}}$ with energy has also been observed in other HBL objects \citep{Aleksic}, supporting the general behavior seen in this class of blazars. In HBLs, the synchrotron peak typically lies in the X-ray band and shifts to higher energies as the source becomes brighter. As a result, X-ray fluxes above the synchrotron peak vary more rapidly than those at lower energies. This behavior, together with the observed $F_{\mathrm{var}}$ trend, suggests that the observed variability is strongly influenced by changes in the high-energy end of the electron distribution, possibly driven by acceleration or cooling processes operating on short timescales. Finally, based on the observed variability, we identified six distinct flux states labeled as VHE, F1, F2, F3, F4, and F5, each marked by shaded vertical bands in Figure~\ref{MWLC}. These states were determined using the Bayesian Block algorithm \citep{Scargle2013} applied to the 3-day binned Fermi-LAT light curve, this method adaptively partitions the light curve into statistically significant segments each representing a time interval with a constant flux level. The selected flux states correspond to portions of the blocks for which simultaneous observations in the X-ray and optical bands are available.


\section{BROAD-BAND SPECTRAL ANALYSIS}\label{broadband}
\indent
To gain insight into the physical mechanisms driving flux variability in 1ES \,1959+650 during its active period, we performed a detailed spectral analysis across different flux states. These states labeled F1, F2, F3, F4, F5 were selected from the multiwavelength light curves based on the availability of simultaneous observations across multiple energy bands and the presence of distinct flux levels. In addition, we defined two states corresponding to VHE $\gamma$-ray detection, labeled VHE-FX1 and VHE-FX2. VHE-FX1 includes simultaneous VHE and $\gamma$-ray data, with optical/UV and X-ray data taken from the F1 state. Similarly, VHE-FX2 includes simultaneous VHE and $\gamma$-ray data, with optical/UV and X-ray data taken from the F2 state. This approach allowed us to construct broadband SEDs for the VHE detections and to investigate whether the X-ray/optical-UV emission during VHE detection more closely resembles the F1 or F2 state.
Within each flux state, we analyzed the spectral behavior in both the $\gamma$-ray and X-ray regimes. The VHE spectrum was calculated at 2 TeV, assuming it to be 0.5 crab units above 1 TeV, as reported by the LHAASO observations. The calculated VHE flux was corrected for extragalactic background light (EBL) absorption.
The $\gamma$-ray spectra were fitted using both log-parabola (LP) and power-law (PL) model, described by the following functional forms:
\begin{equation}
LP: \hspace{1cm}\frac{dN}{dE}
=N_{o}\left(\frac{E}{E_{o}}\right)^{-\alpha-\beta log \frac{E}{E_{o}}}\,
\end{equation}
where, $N_{0}$ denotes the normalization, $\alpha$ represents the photon index at the pivot energy, $E_{0}$, which is fixed at 1732.820 MeV, and $\beta$ quantifies the curvature of the spectrum.
\begin{equation}
PL: \hspace{1cm}\frac{dN}{dE_{0}}
=N_{o}\left(\frac{E}{E_{o}}\right)^{-\Gamma}
\end{equation}
Here, $N_{0}$ represents the normalization, and $\Gamma$ denotes the PL index.
These empirical models are commonly used in blazar studies to characterize curved or hardening/softening spectra (e.g., \citep{2004A&A...422..103M}.
To quantify the significance of spectral curvature in the $\gamma$-ray spectra of the selected flux states, we performed a curvature test using the statistic $TS_{\rm curve} = 2\left[\log \mathcal{L}(\text{LP}) - \log \mathcal{L}(\text{PL})\right]$ \citep{2012ApJS..199...31N}. A curvature is considered statistically significant if $TS_{\rm curve} > 16$.
The computed $TS_{\rm curve}$ values indicate no significant curvature in the $\gamma$-ray spectra for any of the selected flux states (see Table~\ref{TStable}). Consequently, the $\gamma$-ray SED points for all flux states were derived from the power-law (PL) fits.
\vspace{0.2em}

During the interval MJD \,60310 – 60603, \emph{Swift} performed 22 observations.
To generate the X-ray spectrum, the xselect tool was used to extract source and background files for each observation ID. The ancillary response function (ARF) was created using the ``xrtmkarf'' tool, and the grppha task was employed to ensure a minimum of 20 counts per bin. The resulting spectra, corresponding to different flux states, were analyzed with the X-ray spectral fitting software XSPEC, employing Tbabs$\times$powerlaw/log-parabola/BPL models. To account for Galactic absorption, the hydrogen column density for the X-ray observations is fixed at $N_\mathrm{H} = 1.01 \times 10^{21}\,\mathrm{cm^{-2}}$ \citep{Kalberla_2005}. Among the tested models, only the log-parabola model yields statistically acceptable fits to the X-ray spectra with reduced $\boldsymbol{\chi^2}$ values close to unity,  suggesting the presence of intrinsic spectral curvature. 
This curvature is typically attributed to the interplay between particle acceleration and radiative cooling processes \citep{2004A&A...422..103M}.While fitting the X-ray spectrum with the $tbabs \times \text{logparabola}$ model, we found that fixing $N_H$ to this Galactic value yields reduced $\chi^2$ values close to unity in all selected flux states. This indicates that no additional intrinsic absorption component was required. The simple power-law and broken power-law models result in higher $\boldsymbol{\chi^2}$ ($> 2$), indicating that they do not adequately describe the observed spectral curvature. 
The best-fit parameters for all models are summarized in Table \ref{tab:tbabs_model_logparabola} for comparison. For \emph{Swift}-UVOT analysis, the uvotimsum tool was utilized to combine images from individual filters corresponding to the selected flux state. Subsequently, the uvotproduct task was used to extract the flux values for each filter. We noted that the optical/UV spectrum is not well represented by power law model. This implies that the optical/UV flux might contain other emission components apart from the jet emission. Due to negligible errors in these flux values, the broad-band spectral fitting is primarily influenced by the emission in this energy range. To prevent this bias, we introduced additional systematic errors to the optical data, allowing it to be more accurately modeled by a simple power law (reduced $\chi^2 \sim 1$). For states F1, F2, VHE-FX1, VHE-FX2, F3, F4, and F5, these systematic errors were set at 7\%, 7\%, 8\%, 10\%, 9\%, 8\%, and 7\% respectively. The ASCII data, which included the corrected X-ray, optical/UV, and $\gamma$-ray fluxes, were subsequently converted into a Pulse Height Analyser (PHA) file using the HEASARC (High Energy Astrophysics Science Archive Research Center) tool, $ftflx2xsp$.

\vspace{0.2em}

We modeled the broadband SED during different flux states by performing spectral fits using a one-zone leptonic emission model \citep{Shah-2019, Shah-2021}. In this model, we assume that the emission originates from a spherical blob of radius $R$ moving at a relativistic velocity along the jet with a bulk Lorentz factor $\Gamma_{b}$ and at a small angle $\theta$ relative to the observer's line of sight. This relativistic motion, combined with the small viewing angle, leads to Doppler boosting of the observed flux, characterized by the beaming factor $\rm \delta=[\Gamma_{b}(1-\beta\,cos\theta)]^{-1}$, where $\beta$ is the speed of emission region in units of c. The emission region is assumed to be populated with non-thermal electrons, following a broken power-law electron distribution:

\begin{equation}
n(\gamma)d\gamma = 
\begin{cases}
K \gamma^{-p} d\gamma, & \gamma_{\min} < \gamma < \gamma_b \\
K \gamma_b^{q-p} \gamma^{-q} d\gamma, & \gamma_b < \gamma < \gamma_{\max}
\end{cases}
\quad \text{cm}^{-3}
\label{equation(6)}
\end{equation}

where $\gamma$ is the electron Lorentz factor (dimensionless energy), $\gamma_b$ is the break energy, $K$ is the normalization, $p$ and $q$ are the particle spectral indices before and after the break energy. These electrons radiate via synchrotron emission and IC scattering. Given that 1ES\,1959+650 is a HBL object, its broadband SED can typically be explained by synchrotron and SSC processes alone. Accordingly, we consider synchrotron photons produced within the jet as the seed photons for the IC scattering, leading to SSC as the dominant high-energy emission mechanism.
We introduce a new variable $\xi$ such that $\xi =\gamma\sqrt{\mathbb{C}}$, where $\rm \mathbb{C} =1.36\times 10^{-11}\delta B/(1+z)$, to express the electron Lorentz factor $\gamma$. Based on the approach outlined by \citet{Begelman-1984}, the synchrotron flux at energy $\epsilon$ can be determined using the equation
 
 \begin{equation}\label{eq:syn_flux}
 F_{syn}(\epsilon)=\frac{\delta^3(1+z)}{d_L^2} V  \mathbb{A}  \int_{\xi_{min}}^{\xi_{max}} f(\epsilon/\xi^2)n(\xi)d\xi,
 \end{equation}
 
 where, $\rm d_L$  is luminosity distance, $V$ is volume of emission region, $\rm \mathbb A = \frac{\sqrt{3}\pi e^3 B}{16m_e c^2 \sqrt{\mathbb{C}}}$, $\xi_{min}$ and $\xi_{max}$ correspond to the minimum and maximum energy of electron, and f(x) is the synchrotron emmisivity function \citep{Rybicki-1986}.
 The observed SSC flux at energy $\epsilon$ is given by the following equation 
 \begin{equation}\label{eq:ssc_flux}
  \begin{split}
 F_{ssc}(\epsilon) =\frac{\delta^3(1+z)}{d_L^2} V  \mathbb{B} \epsilon & \int_{\xi_{min}}^{\xi_{max}} \frac{1}{\xi^2}  \int_{x_1}^{x_2}   \frac{I_{syn}(\epsilon_i)}{\epsilon_i^2}  \\
&  f(\epsilon_i, \epsilon, \xi/\sqrt{\mathbb{C}}) d\epsilon_i   n(\xi)d\xi
 \end{split}
 \end{equation}
where, $\rm \epsilon_i$ is incident photon energy, $\rm \mathbb{B} = \frac{3}{4}\sigma_T\sqrt{\mathbb{C}}$,  $\rm I_{syn}(\epsilon_i)$ is the synchrotron intensity,  $\rm x_1=\frac{\mathbb{C} \, \epsilon}{4\xi^2(1-\sqrt{\mathbb{C}} \,\epsilon/\xi m_ec^2)}$,  $\rm x_2=\frac{\epsilon}{(1-\sqrt{\mathbb{C}}\,\epsilon/\xi m_e c^2)}$ and

\begin{equation}
f(\epsilon_i, \epsilon, \xi)= 2q\log q+ (1+2q)(1-q)+\frac{\kappa^2q^2(1-q)}{2(1+\kappa q)} \nonumber
\end{equation}
here $\rm q=\frac{\mathbb{C}\epsilon}{4\xi^2\epsilon_i(1-\sqrt{\mathbb{C}}\epsilon/\xi m_ec^2)}$ and $\rm \kappa=\frac{4\xi\epsilon_i}{\sqrt{\mathbb{C}} m_e c^2}$.

 
 \vspace{0.5em}
 
 We numerically solved Equations \ref{eq:syn_flux} and \ref{eq:ssc_flux},  and incorporated the resulting code as a local convolution model in XSPEC for statistical fitting of broadband SEDs. This convolution code provides the flexibility to model the broadband spectrum for any particle energy distribution $n(\xi)$. 
 Under these emission processes, the observed spectrum is mainly governed by 10 parameters like $\xi_{b}$, $\xi_{min}$, $\xi_{max}$, $p$, $q$, $B$, $R$, $\Gamma_{b}$, $\theta$, and norm $N$. The code also enables fitting the SED with jet power ($P_{jet}$) as a free parameter; however, in this case, N must be fixed. In order to reduce the number of free parameters, we adapt a minimalistic emission model. We performed the fitting by allowing the parameters $p$, $q$, $\Gamma_b$, and $B$ to vary freely, while the remaining parameters were fixed at their typical values inferred from the observed broadband spectrum. The fixed parameters were chosen because their values are not well constrained by the available optical/UV, X-ray, and $\gamma$-ray data. Specifically, the jet inclination angle was fixed at $\theta = 2^{\circ}$, the emission region radius at $R = 7.94 \times 10^{16}~\mathrm{cm}$, and the electron energies were set to $\xi_b = 1.1 - 1.6$ (in units of $\sqrt{\mathrm{keV}}$), $\xi_{\mathrm{min}} = (1 - 52) \times 10^{-6}$ (in units of $\sqrt{\mathrm{keV}}$), and $\xi_{\mathrm{max}} = (22 - 25)$ (in units of $\sqrt{\mathrm{keV}}$). The fixed parameter values for different flux states were determined iteratively until the model satisfactorily reproduced the observed spectrum. Due to limited observational coverage across energy bands, uncertainties were estimated only for the four free parameters, as freeing additional parameters caused the XSPEC fitting to fail to converge. The best-fit parameter values, their corresponding reduced $\chi^2$, and the fixed parameters used during the modeling are summarized in Table~\ref{tab:parameters_combined}.

\begin{figure*}
\centering
\begin{subfigure}[b]{0.45\linewidth}
\centering
\includegraphics[scale=0.30,angle=-90]{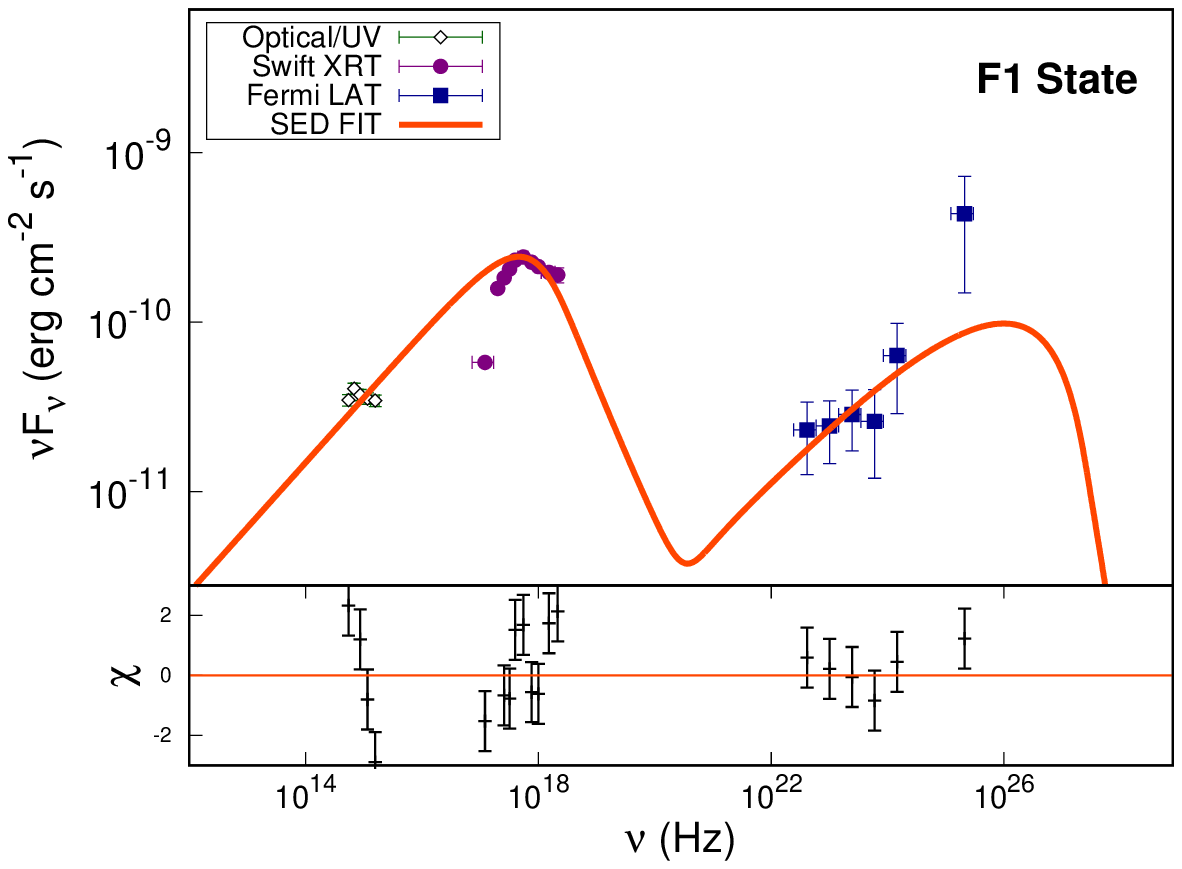}
\vspace{0.25cm}
\end{subfigure}
\vspace{0.2em} 
\begin{subfigure}[b]{0.45\linewidth}
\centering
\includegraphics[scale=0.30,angle=-90]{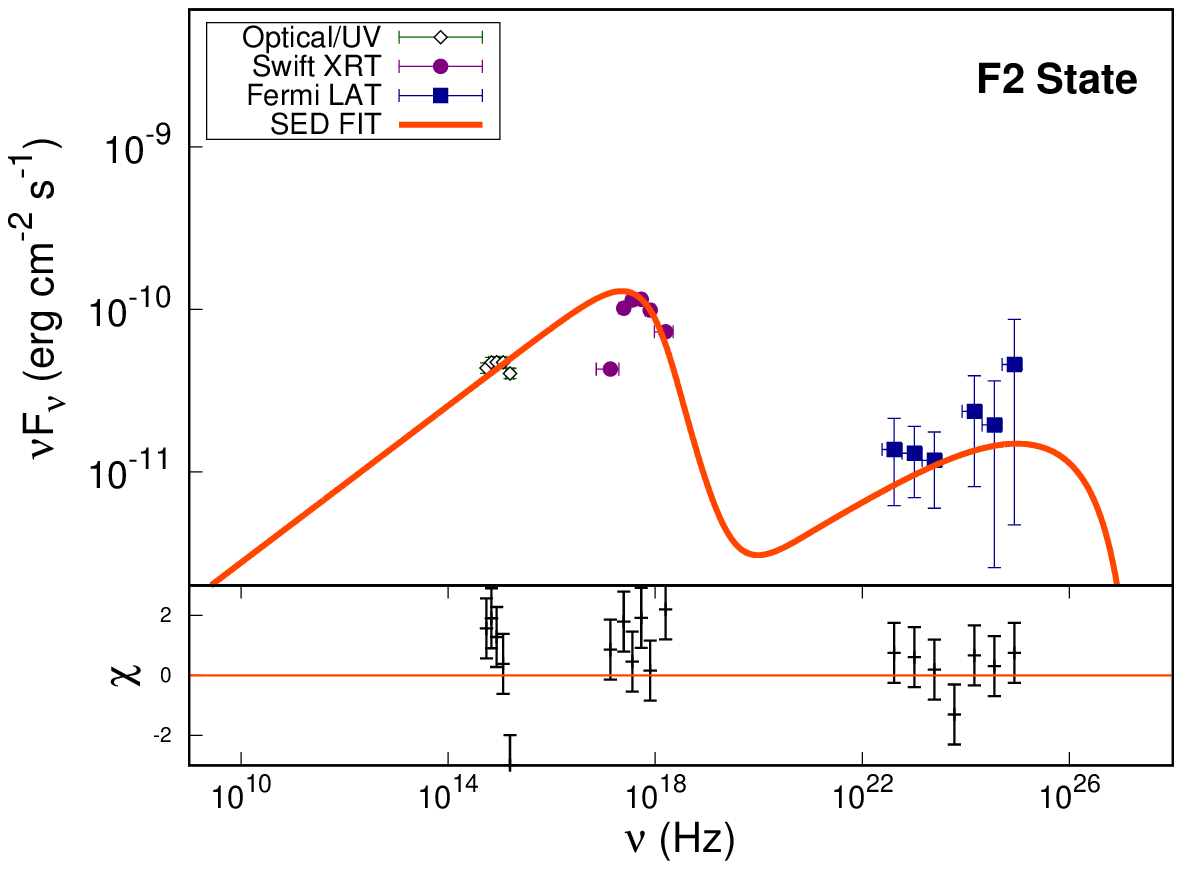}
\vspace{0.25cm}
\end{subfigure}
\vspace{0.2em} 
\begin{subfigure}[b]{0.45\linewidth}
\centering
\includegraphics[scale=0.30,angle=-90]{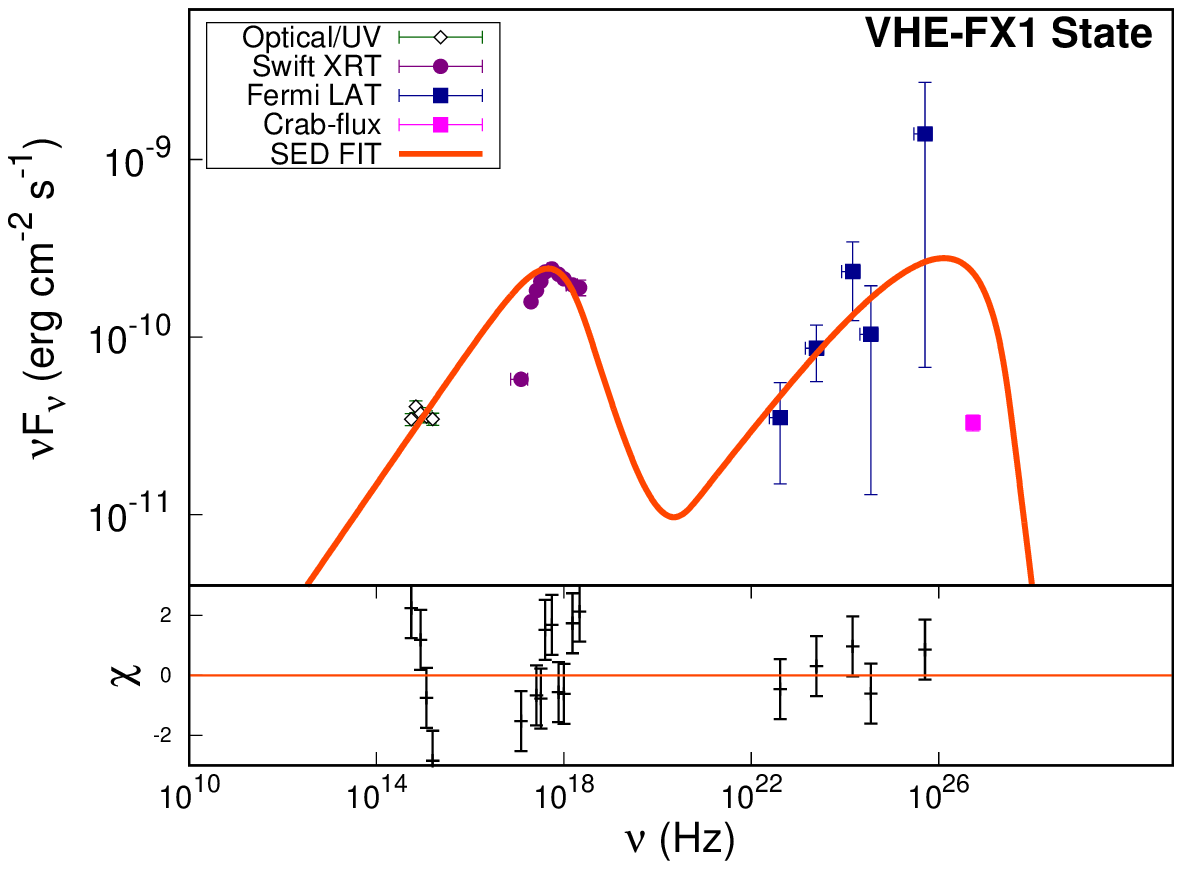}
\vspace{0.25cm}
\end{subfigure}
\vspace{0.2em} 
\begin{subfigure}[b]{0.45\linewidth}
\centering
\includegraphics[scale=0.30,angle=-90]{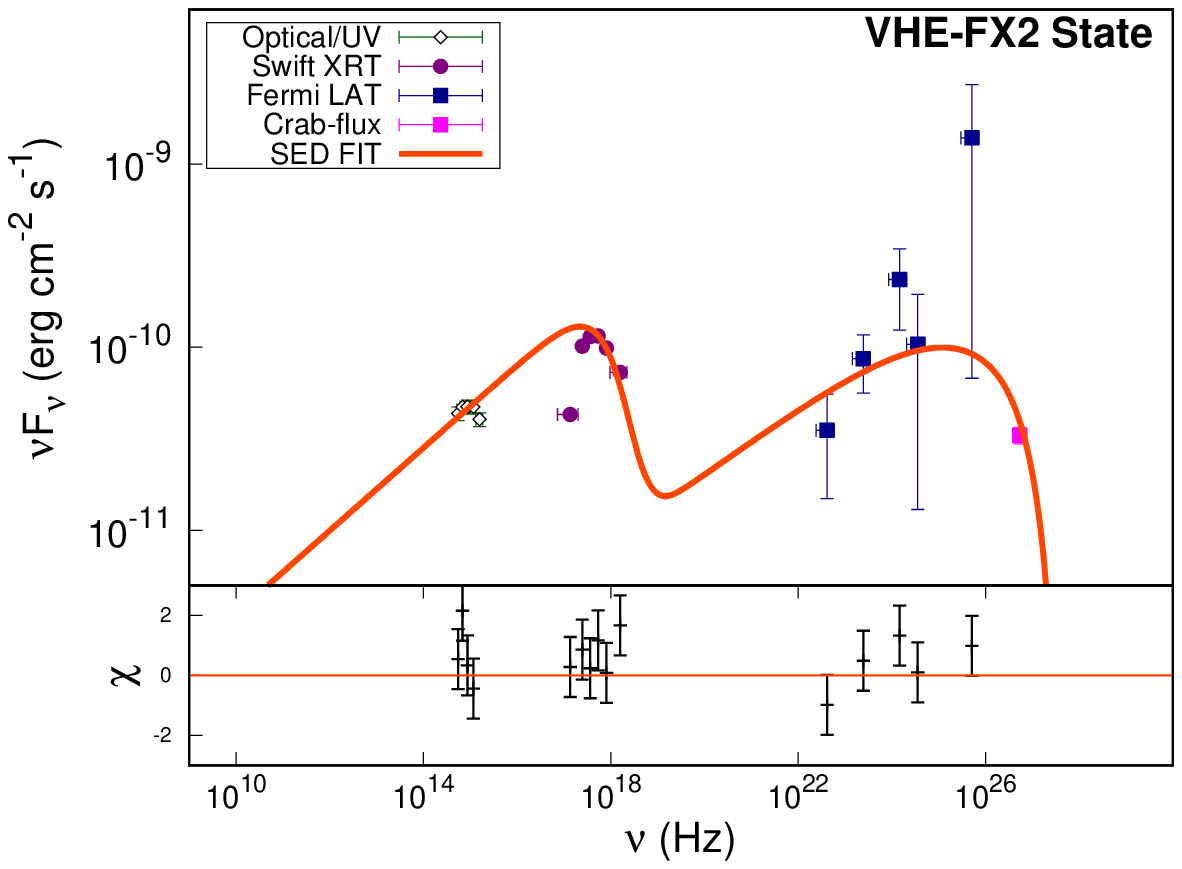}
\vspace{0.25cm}
\end{subfigure}
\vspace{0.2em} 
\begin{subfigure}[b]{0.45\linewidth}
\centering
\includegraphics[scale=0.30,angle=-90]{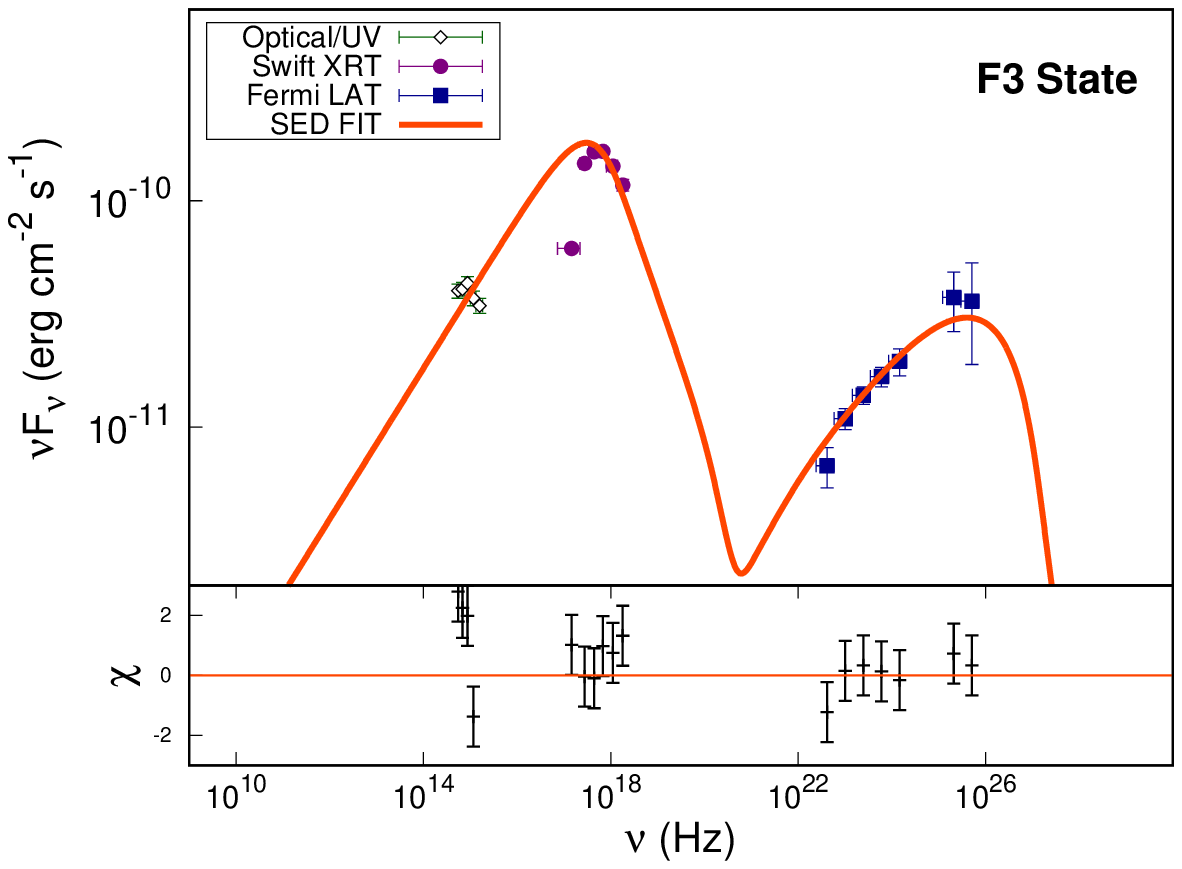}
\vspace{0.25cm}
\end{subfigure}
\vspace{0.2em}
\begin{subfigure}[b]{0.45\linewidth}
\centering
\includegraphics[scale=0.30,angle=-90]{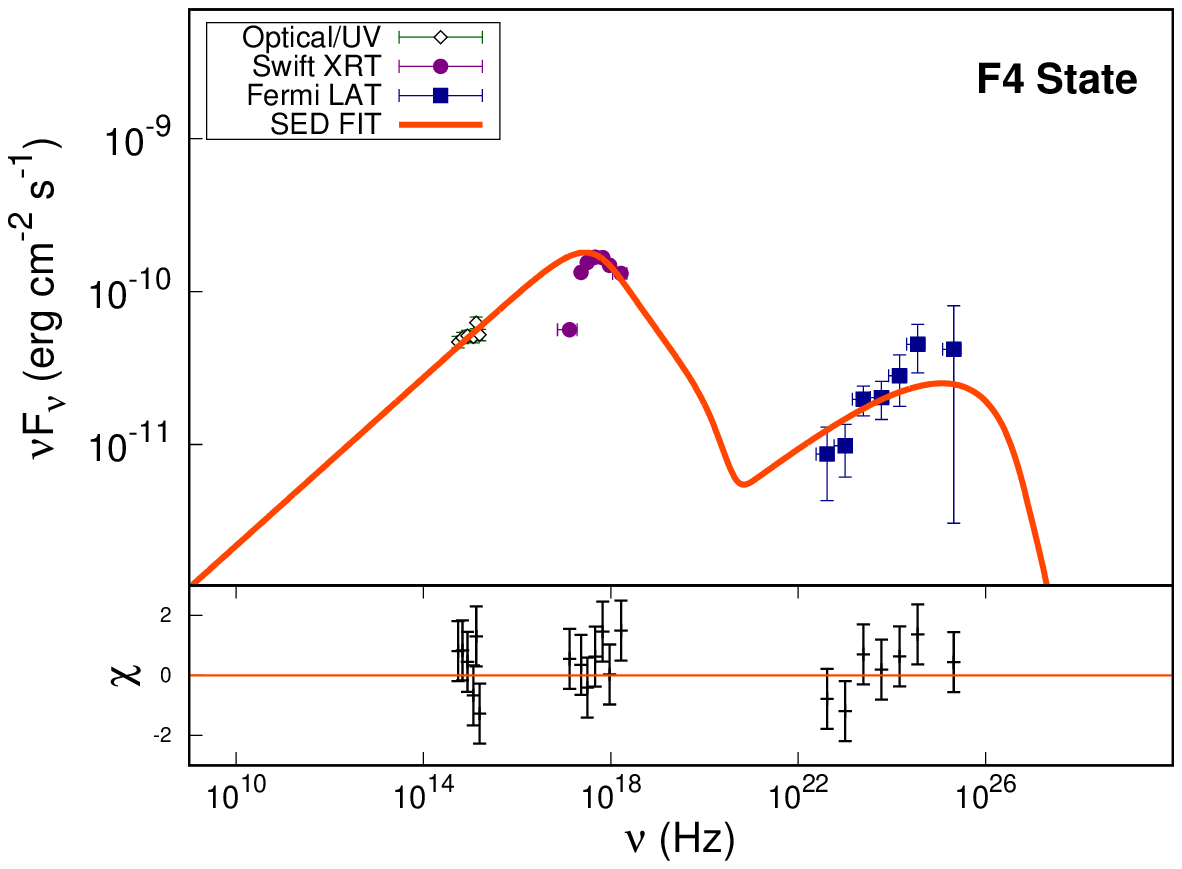}
\vspace{0.25cm}
\end{subfigure}
\vspace{0.2em} 
\begin{subfigure}[b]{0.45\linewidth}
\centering
\includegraphics[scale=0.30,angle=-90]{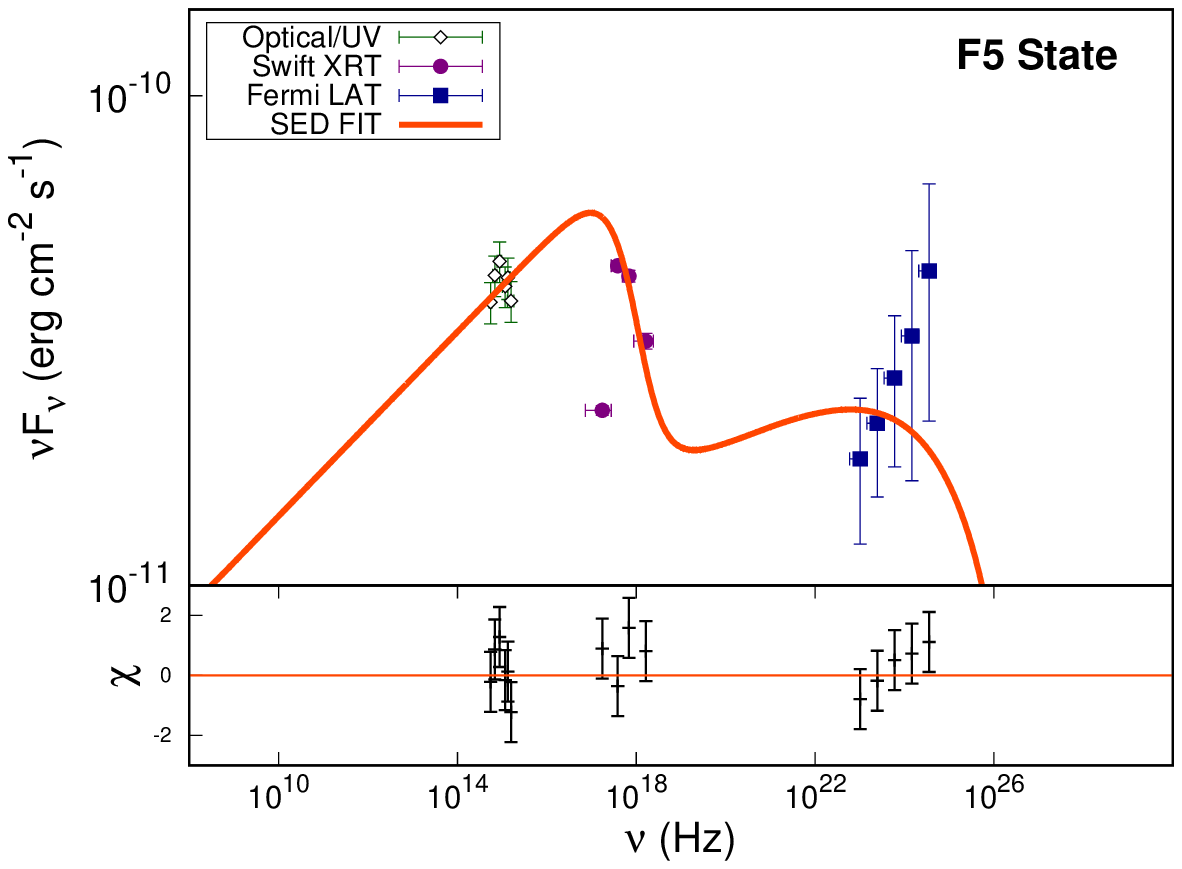}
\vspace{0.25cm}
\end{subfigure}
\caption{Broadband SEDs of 1ES\,1959+650 obtained during different flux states. Top left panel: F1 state; top right panel: F2 state; upper middle left panel: VHE--FX1 state; upper middle right panel: VHE--FX2 state; lower middle left panel: F3 state; lower middle right panel: F4 state; bottom panel: F5 state. In all panels, the flux points denoted by diamond markers correspond to \textit{Swift}-UVOT, circular markers to \textit{Swift}-XRT, blue square markers to \textit{Fermi}-LAT, and pink square markers in the VHE--FX1 and VHE--FX2 panels indicate the Crab flux point (0.5 Crab units above 1~TeV). The solid red curve represents the best-fit model spectrum.}
\label{fig:all_seds}
\end{figure*}

  The spectral fit for all the chosen states are shown in Figure \ref{fig:all_seds}. The kinetic power of the jet can be estimated using $\rm P_{jet} = \pi\,R^2\,\Gamma_{b}^2\,\beta\,c\,(U_{e}+U_{p}+U_{B})$, where, $U_{p}$, $U_{e}$, and $U_{B}$ are the energy densities of protons, electrons, and the magnetic field \citep{Celotti-2008}. In this model, protons are assumed to be cold and do not contribute to radiative processes, consistent with the leptonic scenario. Additionally, the number of protons is assumed to be equal to the number of non-thermal electrons, reflecting a heavy jet structure.
The calculated jet power during different flux states are given in Table \ref{tab:parameters_combined}, and we found that jet power is maximum in the low flux state as compared to the high flux states.

\begin{table*}  
    \centering  
    \renewcommand{\arraystretch}{1.2} 
    \caption{Best-fit spectral parameters for 1ES~1959+650 obtained by fitting the $\gamma$-ray spectrum with power-law (PL) and log-parabola (LP) models across different flux states during the period MJD,60310–60603. The details of Columns are:-  1: Identified flux states;  2: Corresponding time intervals (in MJD);   3: Applied spectral model (PL or LP);   4: Test Statistic (TS);   5: Negative log-likelihood value;   6: Curvature significance.}
     
    \label{TStable}  
    \begin{tabular}{lccccc} 
        \hline  
        State & Period & Model & TS & ${-\log \mathcal{L}}$  & $TS_{curve}$ \\  
        \hline 
        F1 State & 60354\,- 60357 & log\,- parabola & 95.635 & 2417.994 & 0.112 \\  
        -        &      -     & power\,- law     & 95.297 & 2418.050 & -  \\  
        F2 State & 60362\,- 60370 & log\,- parabola & 68.415 &  6758.216 & 0.2199  \\  
        -        &      -     & power\,- law     & 68.193 &  6758.326 & -    \\ 
        F3 State & 60372\,- 60381 & log\,- parabola & 171.341 & 8555.381 & 1.886    \\
        -        &     -       & power\,- law    & 169.708 & 8556.324 & -     \\
        F4 State & 60432\,- 60447 & log\,- parabola & 270.850 & 13859.609 & 2.247     \\
        -        &     -       & power\,- law    & 269.112 & 13860.733 & -      \\
        F5 State & 60510\,- 60532 & log\,- parabola & 76.732 & 16485.200 & 0.184    \\
        -        &     -       & power\,- law    &  76.415 & 16485.384 & -    \\
        \hline  
    \end{tabular}  
\end{table*}

   
\begin{table*}
    \centering
    \renewcommand{\arraystretch}{1.8} 
    \setlength{\tabcolsep}{6pt} 
    \caption{Best-fit X-ray spectral parameters for 1ES\,1959+650 in different flux states using the log-parabola model. Only statistically acceptable fits are shown. Columns: 1: flux state; 2: normalization $(\text{photons}~\mathrm{cm}^{-2}~\mathrm{s}^{-1}~\mathrm{keV}^{-1})$; 3: spectral index at pivot energy 1 keV; 4: curvature parameter; 5: reduced $\chi^2$.}
    \label{tab:tbabs_model_logparabola}
    \begin{tabular}{lcccc}
        \hline
        \textbf{State} & $\mathit{N}\times 10^{-2}$ & $\alpha$ & $\beta$ & $\chi^{2}/{\rm dof}$ \\
        \hline
        F1      & $14.00_{-0.01}^{+0.01}$ & $1.81_{-0.02}^{+0.02}$ & $0.40_{-0.03}^{+0.03}$ & 634.53/515 \\
        F2      & $8.25_{-0.09}^{+0.09}$  & $1.99_{-0.02}^{+0.02}$ & $0.50_{-0.04}^{+0.04}$ & 418.37/395 \\
        VHE--FX1 & $14.00_{-0.01}^{+0.01}$ & $1.81_{-0.02}^{+0.02}$ & $0.40_{-0.03}^{+0.03}$ & 634.53/514 \\
        VHE--FX2 & $8.25_{-0.10}^{+0.10}$ & $1.99_{-0.21}^{+0.21}$ & $0.50_{-0.04}^{+0.04}$ & 481.37/395 \\
        F3      & $11.30_{-0.01}^{+0.01}$ & $1.93_{-0.02}^{+0.02}$ & $0.39_{-0.04}^{+0.04}$ & 447.57/392 \\
        F4      & $11.00_{-0.01}^{+0.01}$ & $1.96_{-0.02}^{+0.02}$ & $0.13_{-0.04}^{+0.04}$ & 485.53/449 \\
        F5      & $3.40_{-0.05}^{+0.05}$  & $2.14_{-0.03}^{+0.03}$ & $0.28_{-0.06}^{+0.06}$ & 307.35/317 \\
        \hline
    \end{tabular}
\end{table*}


\renewcommand{\arraystretch}{2.5} 
\begin{table*}
\centering
\caption{
Best-fit model parameters for 1ES\,1959+650 obtained by fitting the broadband SED model to F1, F2, VHE-FX1, VHE-FX2, F3, F4 and F5 states. The table is organized into three sections: 
The top section lists the free parameters varied during the fit namely: Bulk Lorentz factor of the electron energy distribution ($\Gamma_{b}$); the magnetic field strength ($B$), expressed in units of $10^{-2}$~G; and broken power law indices of the electron distribution before and after the break energy, $p$ and $q$. 
The middle section includes fixed parameters used in the modeling: $\xi_{\rm b}$, $\xi_{\min}$ (in units of $10^{-6}$), and $\xi_{\max}$ represent the break, minimum, and maximum electron energies, respectively, with all energies expressed in units of $\sqrt{\mathrm{keV}}$. The emission region radius ($R = 7.94 \times 10^{16}~\mathrm{cm}$) and jet inclination angle ($\theta = 2^{\circ}$) were kept fixed for all flux states. 
The bottom section presents the logarithm of the jet power ($\log \rm P_{jet}$ (erg s$^{-1})$) and the reduced $\chi$-square value ($\chi^{2}/\mathrm{dof}$) from the spectral fit.}

\label{tab:parameters_combined}
\begin{tabular}{lccccccc}
\hline
\textbf{Parameter} & \textbf{F1} & \textbf{F2} & \textbf{VHE - FX1} & \textbf{VHE - FX2} & \textbf{F3} & \textbf{F4} & \textbf{F5} \\
\hline
\multicolumn{8}{l}{\textit{Free parameters}} \\
$\Gamma_{b}$ & $18.87_{-2.37}^{+3.25}$ & $16.28_{-3.22}^{+4.32}$ & $18.40_{-3.28}^{+4.49}$ & $17.29_{-2.67}^{+3.99}$ & $17.75_{-1.00}^{+1.14}$ & $14.62_{-2.12}^{+2.54}$ & $10.19_{-2.12}^{+2.79}$ \\
$B $ & $0.89_{-0.01}^{+0.01}$ & $2.70_{-0.02}^{+0.01}$ & $0.51_{-0.01}^{+0.01}$ & $0.90_{-0.01}^{+0.02}$ & $1.6_{-0.02}^{+0.02}$ & $2.80_{-0.60}^{+0.50}$ & $3.30_{-0.90}^{+1.40}$ \\
$p$ & $2.23_{-0.02}^{+0.02}$ & $2.51_{-0.02}^{+0.02}$ & $2.23_{-0.02}^{+0.02}$ & $2.55_{-0.01}^{+0.01}$ & $2.33_{-0.03}^{+0.02}$ & $2.45_{-0.03}^{+0.02}$ & $2.82_{-0.03}^{+0.03}$ \\
$q$ & $4.69_{-0.15}^{+0.16}$ & $5.47_{-0.20}^{+0.22}$ & $4.70_{-0.15}^{+0.16}$ & $6.16_{-0.33}^{+0.39}$ & $4.21_{-0.10}^{+0.11}$ & $3.89_{-0.07}^{+0.08}$ & $4.93_{-0.36}^{+0.05}$ \\
\hline
\multicolumn{8}{l}{\textit{Fixed parameters}} \\
$\xi_{b}$ & 1.616 & 1.420 & 1.619 & 1.480 & 1.259 & 1.193 & 1.107 \\
$\xi_{\min}$ & 52.4 & 1.00 & 1.00 & 3.02 & 2.33 & 6.70 & 8.25 \\
$\xi_{\max}$ & 22 & 23 & 22 & 25 & 24 & 23 & 24 \\
\hline
$\rm \log{P_{jet}}$ & 45.71 & 48.66 & 48.03 & 48.56 & 45.54 & 47.21 & 48.27 \\
$\chi^{2}/{\rm dof}$ & 664.40/525 & 538.59/406 & 673.01/525 & 522.46/407 & 466.63/403 & 501.11/461 & 298.37/328 \\
\hline
\end{tabular}
\end{table*}

\section{Summary and Discussion}\label{summary}
In this study, we conducted a comprehensive multi-wavelength temporal and spectral analysis of the VHE blazar 1ES 1959+650 using data from \emph{Fermi}-LAT, \emph{Swift}-XRT, and \emph{Swift}-UVOT over the period MJD\,60310 – 60603. During this interval, the source displayed significant variability across all observed energy bands. The one-day binned $\gamma$-ray light curve ($E > 100$ MeV) exhibited a peak flux of $\rm (5.48 \pm 1.90) \times 10^{-7}\,ph\,cm^{-2}\,s^{-1}$, which is approximately twenty times higher than the average flux reported in the fourth \emph{Fermi}-LAT source catalog \citep{2020yCat..22470033A}. 
The temporal variability analysis reveals a distinct energy dependence, with more pronounced flux fluctuations at higher energies. The correlated variability between the X-ray and $\gamma$-ray bands suggests a common origin for these emissions, consistent with leptonic models in which synchrotron radiation and IC scattering dominate \citep{B_ttcher_2007}. The comparatively moderate variability in the optical/UV bands may indicate contributions from more stable components of the jet, such as a less variable synchrotron emission.
Our calculation of the fractional variability ($F_{\mathrm{var}}$) confirms this energy-dependent trend, with the highest variability detected in the $\gamma$-ray band, followed by the X-ray and UV/optical bands. This increasing $F_{\mathrm{var}}$ with energy is a characteristic behavior observed in HBLs \citep{Aleksic}, reflecting the rapid acceleration and efficient cooling of high-energy electrons responsible for X-ray and $\gamma$-ray emission. These findings underscore the presence of energy-dependent radiative processes shaping the observed variability patterns. The identification of multiple flux states during this campaign also provides a valuable framework for state resolved spectral modeling and offers insight into the physical conditions governing each activity phase.
\vspace{0.2em}

In the X-ray band, we found that the log-parabola model consistently provided a better fit to the spectrum across all flux states, indicating intrinsic spectral curvature. This curvature is commonly attributed to the interplay between particle acceleration and radiative cooling processes. As demonstrated by \citet{2004A&A...422..103M}, a curved spectrum can arise from energy-dependent acceleration mechanisms, where the probability of acceleration decreases with increasing electron energy. Alternatively, an energy-dependent escape rate of electrons from the acceleration region can also lead to curved photon spectra, as suggested by \citet{10.1093/mnras/sty2003, 10.1093/mnrasl/sly086}.
Furthermore, we observed that the X-ray photon index at the pivot energy of 1 keV hardened significantly from $\Gamma_{\rm X} \sim 2.1\,\pm\,0.03$ in the lowest flux state (F5) to $\Gamma_{\rm X} \sim 1.81\,\pm\,0.02 $ in the highest flux state (F1). These spectral changes, accompanied by an increase in flux from  $\rm (1.14\,\pm\,0.05) \times 10^{-10}$ $\text{erg}~\text{cm}^{-2}~\text{s}^{-1}$ to $\rm (5.89\,\pm\,3.20) \times 10^{-10}$ $\text{erg}~\text{cm}^{-2}~\text{s}^{-1}$, suggest a flattening of the underlying electron energy distribution. This trend indicates that the electron spectral index $p$ approaches the canonical value $p \sim 2.0$, which is theoretically expected from first-order Fermi acceleration at relativistic shocks \citep{2000ApJ...542..235K}.
Such a correlation, where the X-ray flux increases alongside spectral hardening is consistent with shock acceleration models, where an enhanced particle injection rate results in both higher synchrotron output and a harder electron spectrum. These findings further support the role of diffusive shock acceleration as a dominant mechanism shaping the spectral evolution in 1ES 1959+650 during its active states.
\vspace{0.2em}

On MJD\,60347, LHAASO detected a VHE $\gamma$-ray flare from 1ES \,1959+650, marking the onset of enhanced activity in VHE regime \citep{atel16437}. During this detection, \emph{Fermi}-LAT recorded the source in the trailing phase of the flare, see Figure~\ref{MWLC}. For reference, the constant TeV $\gamma$-ray flux from the Crab Nebula is reported as $\rm F_{>1,TeV} = (2.89 \pm 0.23) \times 10^{-11}\,cm^{-2}\,s^{-1}\,TeV^{-1}$ \citep{lhaasocollaboration2021performancelhaasowcdaobservationcrab}. In our analysis, we adopted 0.5 Crab units of this flux level to model the broadband SED above 1 TeV. Given that the VHE activity was confined to the period MJD\,60347 – 60348, our primary objective was to analyze and compare the broadband SEDs from this flare epoch with those from other temporal segments. To achieve this, we applied a standard one-zone synchrotron and SSC model to simultaneous multi-wavelength data, including optical/UV observations from \textit{Swift}-UVOT, X-ray data from \textit{Swift}-XRT, and $\gamma$-ray data from \textit{Fermi}-LAT. This modeling was performed across five different flux states (F1, F2, F3, F4, F5). Among these, F1 corresponds to the highest observed X-ray flux, while F5 represents the lowest flux state. Our analysis found that one-zone synchrotron and SSC model provided a satisfactory fit across all these states. 
For VHE\,–FX2 state, the extrapolated SSC model successfully reproduces the VHE flux without invoking additional components, suggesting that the emission remains consistent with a single-zone origin. However, during the brighter VHE\,–FX1 state, the SSC model significantly overpredicts the VHE flux, implying that the lower-energy emission either does not trace the same evolution as in VHE–FX2 or that an additional emission component is required. These results suggest that the X-ray/optical-UV emission during the VHE detections more closely resembles that of the F2 state than the F1 state. Alternatively,  the success of the SSC model in VHE–FX2 supports the idea that, at modest flux levels, a single-zone leptonic mechanism dominates. In contrast, the failure in VHE–FX1 suggests that more complex conditions, such as a second emitting region, structured jet layers, or even a hadronic contribution, may become relevant \citep{2013ApJ...768...54B,Cerruti_2015}. The later scenario bears resemblance to the 2016 flare reported by \citep{2018A&A...611A..44P}, where a one-zone SSC model failed to explain the broadband SEDs in all but one intermediate state (their “F2”), leading to the proposal of a two-zone leptonic framework. Also it is consistent with prior reports of “orphan” VHE flares in 1ES \,1959+650 which have often been interpreted within the context of multi-zone or hadronic models \citep{2004ApJ...601..151K,2005ApJ...621..176B}.
\vspace{0.2em}

The best-fit SED model parameters exhibit systematic trends across the different flux states of 1ES \,1959+650. In higher flux states, we find an increase in the $\Gamma_b$ and a corresponding decrease in the B, along with harder injected electron spectra i.e., lower values of the electron indices $p$ and $q$ during high flux states. Physically, an increase in $\Gamma_b$ implies enhanced Doppler boosting, which is expected during flaring episodes. Simultaneously, a lower B suggests that the emitting region becomes more particle-dominated during these high-activity phases.
The observed hardening of the broken power-law electron distribution reflects more efficient particle acceleration at higher flux states. This ``harder-when-brighter" behavior is a well known trend in blazars \citep{1998A&A...333..452K}, and is consistent with shock acceleration scenarios. Additionally, the trend of increasing Doppler boosting with flux has been interpreted in terms of a fast jet “spine” component becoming dominant during flares, reconciling the observed sub-luminal VLBI speeds with rapid VHE variability \citep{2005A&A...432..401G}. The reduction in B with increasing flux also supports a picture where the jet becomes more kinetic-energy dominated during flaring episodes. These results are broadly consistent with previous studies of HBLs, with 1ES\,1959+650 being a particular example. Interestingly, \citet{2018A&A...611A..44P} found a tight correlation between X-ray and GeV fluxes during the 2016 flare, along with spectral hardening in higher states. Their modeling with a one-zone SSC scenario required extreme parameters such as high Doppler factor ($\delta > 40$), yet still could not explain all flux states without invoking an additional component. Similarly, \citep{2020A&A...638A..14M} reported very hard X-ray/VHE spectra during the same flare, with SSC fits requiring $\delta \sim 50$, suggesting possible contributions from hadronic processes.
In contrast, more recent studies, such as \citet{goswami2023varietyextremeblazarsastrosat}, argue that for typical HBLs, a one-zone SSC model often suffices, and only the most extreme VHE states demand additional components. In our case, the fitted $\Gamma_b$ values remain moderate ($\sim$10\,–\,20), and the one-zone leptonic model satisfactorily explains most flux states including VHE\,–\,FX2.
Moreover, the break energy in the electron distribution ($\xi_b$) shifts from approximately $1.10\,\sqrt{\mathrm{keV}}$ in the low-flux state (F5) to about $1.61\,\sqrt{\mathrm{keV}}$
  in the high-flux state (F1), indicating the presence of higher-energy electrons during brighter phases.
This shift implies not only more efficient particle acceleration but also results in a corresponding shift in the peak frequencies of the emitted synchrotron and IC components. In particular, the synchrotron peak moves to higher energies, often observed as a hardening of the X-ray spectrum. Similarly, the high-energy (IC) peak in the SED may shift toward the GeV, enhancing detectability in the VHE band.
Additionally, a higher break energy often correlates with shorter cooling timescales for electrons, potentially leading to more rapid flux variability. 
The modeled jet power also shows relatively high values ($>10^{46}$ erg s$^{-1}$) across all flux states. Under certain assumptions about jet composition (e.g., presence of cold protons), such high power may suggest a significant energy budget carried by non-radiating particles.
Overall, the results point to a dynamic jet environment, where flaring episodes involve coordinated changes in Doppler boosting, particle acceleration, and magnetization. This behavior is broadly consistent with the blazar sequence. The SED modeling of the VHE–FX1 state highlights the complexity of such flares and aligns with this source’s known history of “orphan” VHE outbursts. Our findings reinforce the view that BL Lac jets are highly dynamic systems, with energy dissipation governed by varying conditions in both the particle and field components of the emission region \citep{2019ApJ...887..133B}.

\section{Data Availability}\label{Data Availability}
The data and the model used in this article will be shared on reasonable request to the corresponding author, Zahir Shah (email: \href{mailto:shahzahir4@gamil.com}{shahzahir4@gmail.com}) or Peer Anjum (email: \href{mailto:peer.anjum@iust.ac.in}{peer.anjum@iust.ac.in}).

\bibliographystyle{mnras}
\bibliography{main_bib} 








\label{lastpage}
\end{document}